\documentclass{article}

\usepackage{arxiv}

\usepackage[utf8]{inputenc} 
\usepackage[T1]{fontenc}    
\usepackage{hyperref}       
\usepackage{url}            
\usepackage{amsmath}
\usepackage{booktabs}       
\usepackage{amsfonts}       
\usepackage{nicefrac}       
\usepackage{microtype}      
\usepackage{cleveref}       
\usepackage{lipsum}         
\usepackage{graphicx}
\usepackage{natbib}
\usepackage{doi}

\title{Scientific Machine Learning for Modeling and Simulating Complex Fluids}

\author{ \href{https://orcid.org/0000-0002-1251-5461}{\includegraphics[scale=0.06]{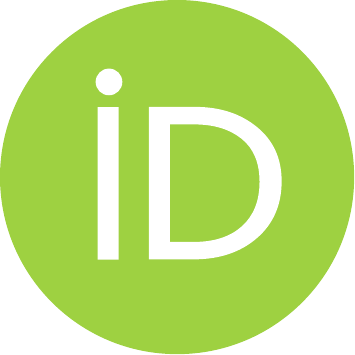}\hspace{1mm}Kyle R.~Lennon} \\
	Department of Chemical Engineering\\
	Massachusetts Institute of Technology\\
	Cambridge, MA 02142 \\
	\texttt{krlennon@mit.edu} \\
	\And
	\href{https://orcid.org/0000-0001-8323-2779}{\includegraphics[scale=0.06]{orcid.pdf}\hspace{1mm}Gareth H.~McKinley} \\
	Department of Mechanical Engineering\\
	Massachusetts Institute of Technology\\
	Cambridge, MA 02139 \\
	\texttt{gareth@mit.edu} \\
	\And
	{James W.~Swan}\thanks{Deceased} \\
	Department of Chemical Engineering\\
	Massachusetts Institute of Technology\\
	Cambridge, MA 02142 \\
}

\renewcommand{\shorttitle}{Rheological Universal Differential Equations}

\hypersetup{
pdftitle={Rheological Universal Differential Equations},
pdfsubject={q-bio.NC, q-bio.QM},
pdfauthor={Kyle R.~Lennon, Gareth H.~McKinley, James W.~Swan},
pdfkeywords={machine learning, rheology, soft matter, constitutive equation},
}

\begin{document}
\maketitle

\begin{abstract}
	The formulation of rheological constitutive equations -- models that relate internal stresses and deformations in complex fluids -- is a critical step in the engineering of systems involving soft materials. While data-driven models provide accessible alternatives to expensive first-principles models and less accurate empirical models in many engineering disciplines, the development of similar models for complex fluids has lagged. The diversity of techniques for characterizing non-Newtonian fluid dynamics creates a challenge for classical machine learning approaches, which require uniformly structured training data. Consequently, early machine learning constitutive equations have not been portable between different deformation protocols or mechanical observables. Here, we present a data-driven framework that resolves such issues, allowing rheologists to construct learnable models that incorporate essential physical information, while remaining agnostic to details regarding particular experimental protocols or flow kinematics. These scientific machine learning models incorporate a universal approximator within a materially objective tensorial constitutive framework. By construction, these models respect physical constraints, such as frame-invariance and tensor symmetry, required by continuum mechanics. We demonstrate that this framework facilitates the rapid discovery of accurate constitutive equations from limited data, and that the learned models may be used to describe more kinematically complex flows. This inherent flexibility admits the application of these `digital fluid twins’ to a range of material systems and engineering problems. We illustrate this flexibility by deploying a trained model within a multidimensional computational fluid dynamics simulation -- a task that is not achievable using any previously developed data-driven rheological equation of state.
\end{abstract}


\section*{Significance Statement}

The development of models that accurately describe the behavior of complex fluids under flow is a longstanding challenge in soft materials engineering. Data-driven models for these fluids have yet to make a significant impact, largely due to their inflexibility -- once trained, they are able to describe only a single experiment. We propose a framework for learning accurate constitutive models that are not fixed to a particular experiment, but may instead describe the fluid in any flow configuration. We demonstrate that these models may be trained on data obtained in a laboratory setting, then applied to predict fluid properties in a multidimensional simulation of an industrially relevant flow. This framework opens new avenues to rapid soft material design and engineering.

\section*{Introduction}

The advent of fast, scalable, and accessible data-driven tools and techniques has revolutionized much of the scientific and engineering landscape \citep{Carleo2019}. Machine learning (ML) methods are now widespread in fields such as computational chemistry \citep{Goh2017}, biomedical imaging \citep{Park2018}, and particle physics \citep{Radovic2018}. The success of ML approaches in many traditionally difficult problems has recently galvanized efforts to leverage ML within scientific simulations. In this young field of `scientific machine learning' \citep{Rackauckas2020}, traditional ML tools often serve as fast surrogate models for steps in the simulation that are computationally burdensome, such as electronic structure calculations in \emph{ab initio} molecular dynamics \citep{Ellis2021}, or as accurate system-specific equations of state when first-principles models are not practical, such as models for the Reynolds stress in turbulent fluid flows \citep{Ling2016}. These fast and accurate machine-learning-enhanced simulations promise to substantially accelerate the process of materials design and engineering.

Despite their success in other fields, ML methods have yet to substantially impact the field of complex fluid rheology, whose broad aim is to characterize the (tensorial) relationship between stresses and deformation within flowing materials. One particular challenge in this field is that rheological data sets are at once too scarce and too diverse to enable traditional ML approaches -- their scarcity a reflection of the time- and material-intensive nature of bulk rheometry, and their diversity a product of the many rheometric protocols and tools used to characterize the mechanical behavior of complex fluids \citep{Tschoegl1989,Bird1987}. The success of data-driven rheometry thus depends on the ability to simultaneously assimilate different types of experimental data in a unified manner, a notable weakness of many common ML approaches. Within the context of simulations of complex fluids, these challenges are exacerbated by the fact that many complex fluids and soft matter systems exhibit a `fading memory' of their deformation history \citep{Bird1987}. In contrast to many other systems, representations that treat the constitutive relationship between stress and deformation as instantaneous (such as feedfoward neural networks \citep{Svozil1997}) are again insufficient.

Early efforts to apply ML to rheological modeling have attempted to overcome the barrier of data scarcity by incorporating physical or empirical knowledge into the framework for model fitting \citep{Mahmoudabadbozchelou2021a,Mahmoudabadbozchelou2021b,Mahmoudabadbozchelou2022a}. For instance, physics-informed neural networks (PINNs) penalize differences between the predictions of the networks and the predictions of available constitutive equations or physical conservation laws during training \citep{Raissi2019}. Although such physical constraints have been shown to improve the performance of neural networks when the amount of training data is limited, these physical and constitutive laws are typically enforced by soft constraints, and therefore the trained networks are not guaranteed to respect these laws for all inputs. Moreover, because these physics-informed neural networks require the user to specify an extant model during training, they have been limited to the problems of model identification or model parameterization \citep{Saadat2022,Thakur2022}, rather than the \emph{discovery} of more accurate equations of state. While it is possible to tune models to closely describe experimental data using a multifidelity approach \citep{Mahmoudabadbozchelou2021a}, the feedforward structure of the final model constrains their application to a single time-varying deformation protocol \citep{Mahmoudabadbozchelou2022a}. Porting the learned model to another time-varying deformation protocol necessitates retraining the model on data that is synthetically generated by a known constitutive model, thus intrinsically limiting the generality of the approach \citep{Mahmoudabadbozchelou2022b}.

To design a framework for learning rheologically invariant models that describe the mechanical response of complex fluids in arbitrary deformations, it is instructive to consider the purpose of constitutive equations in rheology. Constitutive equations for complex fluids are equations of state that provide mathematical relationships between an arbitrary deformation history of a material element and the corresponding tensorial stress felt by that material element at every point in that history \citep{Green1957,Green1959,Oldroyd1984}. The most general of these equations consider deformations and stresses in three dimensions, such that the constitutive equation is not bound to a particular experimental protocol or observable \citep{Oldroyd1950}. Therefore, one may use these constitutive equations, together with the equations of motion, to conduct simulations in different conditions \citep{Oldroyd1984}. The previously described `end-to-end' machine-learning-based constitutive equations, on the other hand, output a specific observable, such as the shear stress, and often rely on structured input data corresponding to specific experiments. These models therefore fail to provide this critical feature of constitutive equations -- namely, the ability to make predictions for kinematic histories beyond the set of experimental observations.

This fundamental limitation of present ML approaches arises because the details of experimental observation are ingrained within their architectures, for instance by selecting a particular experimental observable as the output of a feedforward neural network. This is in contrast to physical constitutive equations that apply to real materials, which exist independently from the process of experimental observation. This realization is the impetus for the novel class of machine learning constitutive equations that we introduce in this work, which we call rheological universal differential equations (RUDEs). These models encapsulate all components of the tensors that describe deformations and stresses in three dimensions, and therefore may be used to describe the response of a material to any deformation history. The relationship between experimental observations and the predictions of this model is contained entirely within the loss function; thus, RUDEs can be trained using any experimental observable. Once trained, a RUDE may similarly be used to predict experimental observables or the response to experimental protocols not included in the training set, including the behavior of a complex fluid in a multidimensional flow.

The ability to train a single RUDE on multiple types of rheometric data is critical to leveraging the diverse data sets characteristic of rheology and thermochemical characterization protocols in soft material science more broadly. It also ensures that no available data need be wasted during training, ameliorating to some extent the issue of data scarcity. As previously discussed, another remedy for scarce data is to incorporate physical or empirical laws within the learning framework. This too is possible with RUDEs. However, unlike the soft constraints used to enforce these laws with end-to-end approaches, these laws may be directly incorporated into the structure of a RUDE; thus, physical laws such as frame invariance are respected by default, allowing RUDEs to converge to highly general models even with limited training data. 

In the following Section, we begin by developing the mathematical framework for RUDEs, after which we present examples using both synthetic and real experimental data, demonstrating that RUDEs are a powerful tool for discovering robust, general, and transferable constitutive equations directly from data. To emphasize these critical features, we conclude by integrating a RUDE trained only on one-dimensional oscillatory shear data into a computational fluid dynamics tool to analyze a complex two-dimensional mixed flow featuring both shearing and extensional kinematic components, thus demonstrating that the model's accuracy extends to more complicated multidimensional flows.

\section*{Rheological Universal Differential Equations}

The salient features of RUDEs described above arise from their origin in scientific machine learning, as an extension of the `universal differential equation' \citep{Rackauckas2020}. These models embed universal approximators -- typically, neural networks -- within differential equations. By defining a RUDE as a coupled system of universal differential equations describing each component of the stress ($\boldsymbol{\sigma}$) and deformation ($\boldsymbol{\dot{\gamma}}$) tensors, we obtain a model that completely describes the mechanical state of a viscoelastic material, and may therefore be used to predict any experimental observable related to deformation or flow.

In addition to their data-agnostic architecture, RUDEs present another substantial advantage over present machine learning constitutive equations. Specifically, one may straightforwardly embed physical or empirical information into the differential equation alongside the neural network, resulting in a machine learning model that automatically satisfies these physical or empirical constraints \citep{Rackauckas2020}. Here, we choose a widely employed constitutive equation that describes fundamental aspects of viscoelasticity, called the Oldroyd-B model \citep{Oldroyd1950}, to structure the RUDEs. The neural network $\boldsymbol{F}(\boldsymbol{\sigma},\boldsymbol{\dot{\gamma}})$ is simply an additional term embedded within this model framework:
\begin{equation}
    \boldsymbol{\sigma} + \tau_1\overset{\nabla}{\boldsymbol{\sigma}} + G_0\boldsymbol{F}(\boldsymbol{\sigma},\boldsymbol{\dot{\gamma}}) = G_0\tau_1\left(\boldsymbol{\dot{\gamma}} + \tau_2\overset{\nabla}{\boldsymbol{\dot{\gamma}}}\right).
    \label{eq:rude}
\end{equation}
Here, $\overset{\nabla}{\boldsymbol{x}}$ represents the upper-convected derivative of the tensor-valued quantity $\boldsymbol{x}$ \citep{Bampi1980}. Because linear and nonlinear viscoelasticity is inherent within the model structure, the neural network may focus exclusively on learning material-specific features, thereby reducing the amount of experimental data required to train an accurate model.

\begin{figure*}
    \centering
    \includegraphics[width=\textwidth]{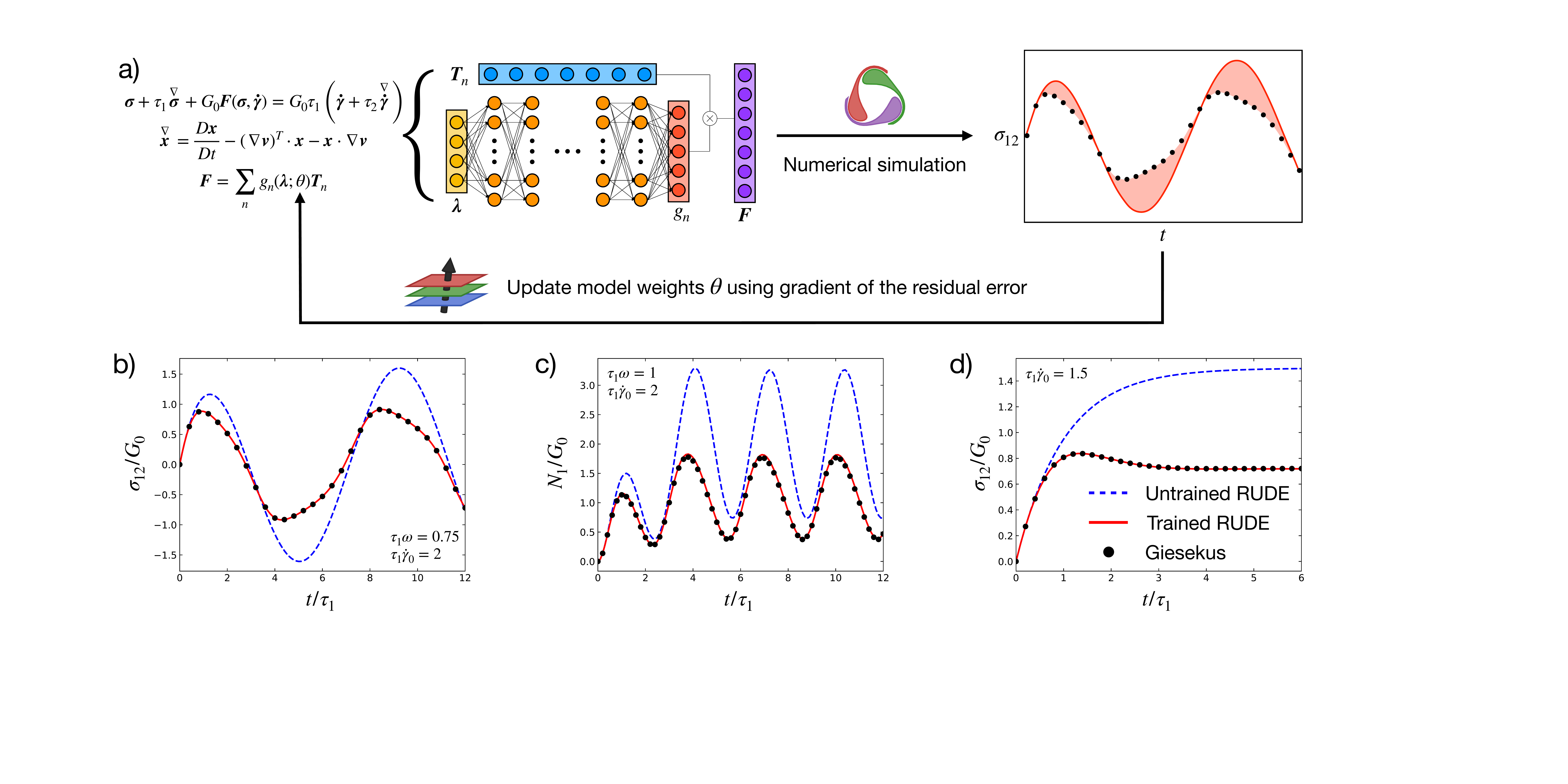}
    \caption{(a) Schematic depiction of a rheological universal differential equation (RUDE) within the training loop. The RUDE comprises a generalized form of Oldroyd's differential tensorial formulation with an embedded tensor basis neural network (TBNN). The TBNN computes an expansion of the arbitrary function $\boldsymbol{F}(\boldsymbol{\sigma},\boldsymbol{\dot{\gamma}})$ in terms of tensor products $\boldsymbol{T}_n$ of the stress, rate-of-strain, and identity tensors, with coefficients determined by the outputs of a neural networks with inputs $\lambda_n$ being the invariants of tensors $\boldsymbol{T}_n$. The RUDE is evolved numerically by a Runge-Kutta method with adaptive time stepping (DifferentialEquations.jl). The mean squared error serves as the loss metric, and gradients of this loss with respect to the TBNN parameters $\theta$ are computed quickly using adjoint sensitivity analysis (Flux.jl). (b)-(d) Example test evaluation of a RUDE trained on a synthetic data set consisting of the time- and rate-dependent shear stress in eight simulated large amplitude oscillatory shear (LAOS) experiments for the Giesekus model (with model parameters $\tau_2 = 0$, $\alpha = 0.8$). Black circles represent the test data (i.e. the simulated response of the Giesekus model), red lines indicate the trained RUDE, and blue lines indicate the RUDE prediction before training. Panel (b) depicts the predicted shear stress response in LAOS at an intermediate frequency not seen in the training set. Panel (c) depicts the predicted normal stress at a frequency seen during training. Panel (d) depicts the predicted shear stress response in a different kinematic scenario corresponding to the startup of steady shear flow.}
    \label{fig:tbnn}
\end{figure*}

We select the Oldroyd-B model in part because it preserves a key physical symmetry for isotropic materials: invariance to a rotating frame of reference \citep{Oldroyd1950}. This physical principle, which Oldroyd termed `rheological invariance', was central in the development of his constitutive framework \citep{Oldroyd1984}. To preserve this physical constraint when incorporating a neural network into the model, we must construct a frame-invariant architecture for that network. Here, we employ a frame-invariant architecture called the tensor basis neural network (TBNN) \citep{Ling2016}. The details of this architecture are presented in the Materials and Methods section, and a schematic depiction is presented in Figure \ref{fig:tbnn}. This architecture maintains frame invariance by treating a polynomial expansion of the arbitrary, tensor-valued function $\boldsymbol{F}(\boldsymbol{\sigma},\boldsymbol{\dot{\gamma}})$ in matrix products $\boldsymbol{T}_n$ of its arguments ($\boldsymbol{\sigma}$ and $\boldsymbol{\dot{\gamma}}$) and the identity tensor. The coefficients of these terms are themselves arbitrary functions of the invariants $\lambda_n$ of the tensors $\boldsymbol{T}_n$, and these coefficients are modeled by the outputs $g_n$ of a fully-connected feedforward neural network:
\begin{equation}
    \boldsymbol{F} = \sum_n g_n(\boldsymbol{\lambda})\boldsymbol{T}_n. \label{eq:tbnn}
\end{equation}
The complete set of $\boldsymbol{T}_n$ and $\lambda_n$ obtained from the two symmetric tensors $\boldsymbol{\sigma}$ and $\boldsymbol{\dot{\gamma}}$ can be computed by repeated application of the Cayley-Hamilton theorem \citep{Spencer1958,Spencer1959}. We list these tensors and invariants in the Materials and Methods section. Note that while we have written the arguments of $\boldsymbol{F}$ as the dimensional stress ($\boldsymbol{\sigma}$) and rate-of-deformation ($\boldsymbol{\dot{\gamma}}$) tensors here for the sake of clarity, the performance of the neural network during training is improved by instead supplying the dimensionless versions of these tensors as arguments, $\boldsymbol{\tilde{\sigma}} = \boldsymbol{\sigma}/G_0$ and $\boldsymbol{\tilde{\dot{\gamma}}} = \tau_1\boldsymbol{\dot{\gamma}}$ where $G_0$ and $\tau_1$ are the characteristic modulus and relaxation time for the viscoelastic material of interest, as defined in equation \ref{eq:rude}.

In previous machine learning approaches for rheological modeling, both the inputs and outputs of the neural networks were directly observed in simulations or experiments, and therefore the gradients with respect to model parameters that are used during model training could be obtained directly by backpropagation through the network. In the case of RUDEs, however, there is no direct supervision signal for many of the inputs and outputs of the TBNN. Therefore, it is necessary to numerically evolve the RUDE according to some training deformation protocol, and evaluate the model loss by comparing the resulting solution to discrete data corresponding to a particular experimental observable. Because RUDEs are fully three-dimensional, the deformation protocol used to obtain the training data, and the particular experimental observables in the training data set, may be chosen freely. For instance, both extensional and shear flows may form the training set for the same RUDE, and the RUDE may be trained on shear stress data, normal stress data, or data describing any combination of elements of the stress tensor. Restricting ourselves for now to shear stress data in simple shear flows, the loss metric may be written as follows:
\begin{equation}
    \mathcal{L} = \sum_{\mathcal{D}_j \in \{\mathcal{D}_j\}}\sum_{(t^{(i)},\sigma_{12}^{(i)}) \in \mathcal{D}_j}\frac{(\sigma_{12}^{(i)} - \hat{\sigma}_{12}^{(i)})^2}{\max_{k} |\sigma_{12}^{(k)}|}, \label{eq:loss}
\end{equation}
where $t^{(i)}$ and $\sigma_{12}^{(i)}$ represent the $i$th time and shear stress data point, $\hat{\sigma}_{12}^{(i)}$ represents the shear stress predicted by the RUDE at the corresponding time point, and $\{\mathcal{D}_j\}$ represents the set of experiments $\mathcal{D}_j$ used in training.


Direct backpropagation from this loss would require knowledge of the integration routine used to obtain each $\hat{\sigma}_{12}^{(i)}$; however, it is not practical to systematically enumerate the backpropogation rules pertaining to each of these routines. Therefore, we instead train the RUDE by numerically evolving a set of adjoint differential equations, which has the same size as the system of differential equations corresponding to a RUDE (i.e. six differential equations) \citep{Cao2003}. These adjoint equations are constructed using automatic differentiation, and are independent of the integrator used to evolve the RUDE. Moreover, this adjoint sensitivity analysis avoids numerically evolving the Jacobian of the stress tensor with respect to the parameters of the TBNN, a much more expensive calculation than the evolution of six adjoint differential equations. This sensitivity analysis is implemented in the Julia language, in particular by the DifferentialEquations.jl, Flux.jl, and DiffEqFlux.jl packages \citep{Rackauckas2017,Rackauckas2019}. Thus, computation of the gradients required for model optimization is done in an efficient and parallelizable manner, and is independent of specific features of the training data (such as its size and time step). This flexibility allows RUDEs to be trained simultaneously on experimental data sets obtained using different flow histories, and with different experimental observables.

\section*{Results}

Because RUDEs relate the elements of the stress and rate-of-strain tensor in the form of a materially objective differential equation, they represent a tensorial frame-invariant constitutive equation capable of describing viscoelastic fluids in arbitrary flows. An immediate application of RUDEs, therefore, is in model discovery. Other machine learning approaches have focused primarily on using neural networks to either accelerate solutions to constitutive equations, or to refine the predictions of a known constitutive equation for specific flows and deformation protocols (e.g. the steady material response in steady simple shear flow) -- resulting in a so-called `digital rheometer twin' \citep{Mahmoudabadbozchelou2022b}. A trained RUDE, however, represents a digital twin for the underlying viscoelastic fluid itself, and this `digital fluid twin' is, in principle, capable of simulating the fluid response in arbitrary rheometric deformations as well as multidimensional complex flows.

In this Section, we demonstrate the application of RUDEs to the problem of model discovery first with the example of reproducing the behavior of a (known) constitutive equation from simulated rheometric data, followed by the discovery of a constitutive equation for a real viscoelastic fluid from experimentally obtained data. Finally, we underscore that a trained RUDE represents a well-posed materially objective constitutive model that can be applied out-of-the-box to complex fluid simulations by incorporating the trained model into an open-source computational fluid dynamics software package, and conducting a simulation in a benchmark multidimensional flow.

\subsection*{Reproducing a Known Constitutive Equation}

Many well-known nonlinear constitutive equations are described exactly by equation \ref{eq:rude} with a specific form for $\boldsymbol{F}(\boldsymbol{\sigma},\boldsymbol{\dot{\gamma}})$ \citep{Lennon2021}. For instance, the Giesekus model has $\boldsymbol{F}(\boldsymbol{\sigma},\boldsymbol{\dot{\gamma}}) = \frac{\alpha}{G_0^2} \boldsymbol{\sigma}\cdot\boldsymbol{\sigma}$ with the mobility parameter $\alpha \in [0,1)$ \citep{Giesekus1982}. A suitable initial test for the RUDE framework is therefore to reproduce the behavior of the Giesekus model from simulated data. To facilitate this test, we generate a training data set comprising eight synthetic large amplitude oscillatory shear (LAOS) experiments for the Giesekus model with $\tau_2 = 0$ and $\alpha = 0.8$ \citep{Hyun2011}. Stress and time are made dimensionless by the parameters $G_0$ and $\tau_1$, so the particular values of these parameters are inconsequential. The LAOS data were generated with shear rate protocols of the form: $\dot{\gamma}(t) = \mathrm{Wi} \cos (\mathrm{De} t)$, with Deborah number $\mathrm{De} \equiv \tau_1\omega \in \{0.33, 0.5, 1, 2\}$ for each Weissenberg number $\mathrm{Wi} \equiv \tau_1\dot{\gamma}_0 \in \{1, 2\}$. Importantly, only the simulated shear stress was recorded for training, even though the nature of the convected Oldroyd derivative in the nonlinear constitutive formulation gives rise naturally to the growth and evolution of oscillating normal stress components to the material response. The RUDE was trained by a continuation procedure, in which experiments were incorporated into the training set successively after epochs of 200 iterations.

Once the RUDE was trained, it was tested on deformation protocols and observables outside of the training set, as depicted in Figure \ref{fig:tbnn}b)-d). First, we demonstrate the ability of the trained RUDE to interpolate within the training data set, by studying a LAOS experiment at a new dimensionless frequency: $\mathrm{De} = 0.75$ (with $\mathrm{Wi} = 2$). From Figure \ref{fig:tbnn}b), it is clear that the trained RUDE (red line) closely approximates the simulated data (black circles), and has substantially improved upon the pre-training model (dashed blue line). Indeed, the difference between the trained model and data is nearly imperceptible, despite the fact that the model did not use this data during training. We next illustrate the transferability of the RUDE to different observables, such as the first normal stress difference, $N_1(t;\mathrm{Wi})$. Figure \ref{fig:tbnn}c) depicts the prediction of the first normal stress difference, $N_1 = (\sigma_{11} - \sigma_{22})$ for a LAOS test with $\mathrm{De} = 1$ and $\mathrm{Wi} = 2$. Although the same deformation conditions were used during training, the training algorithm only employed the shear stress data, and therefore the time- and strain-dependent normal stresses represent data unseen by the model during training. Still, we observe very close agreement between model predictions and the simulated data, and substantial improvement from the pre-training state. Finally, we test the portability of RUDEs to different deformation protocols, with Figure \ref{fig:tbnn}d) depicting predictions in the startup of steady shear flow under nonlinear conditions corresponding to $\mathrm{Wi} = 1.5$. Although only time-periodic LAOS data was used in training, the frame invariance and material objectivity of the trained RUDE enables accurate predictions of the startup response, including the stress overshoot and ultimate steady shear plateau. Notably, neither of these features were evident in any training data, nor were they well-described by the pre-training model.

\begin{figure*}
    \centering
    \includegraphics[width=\textwidth]{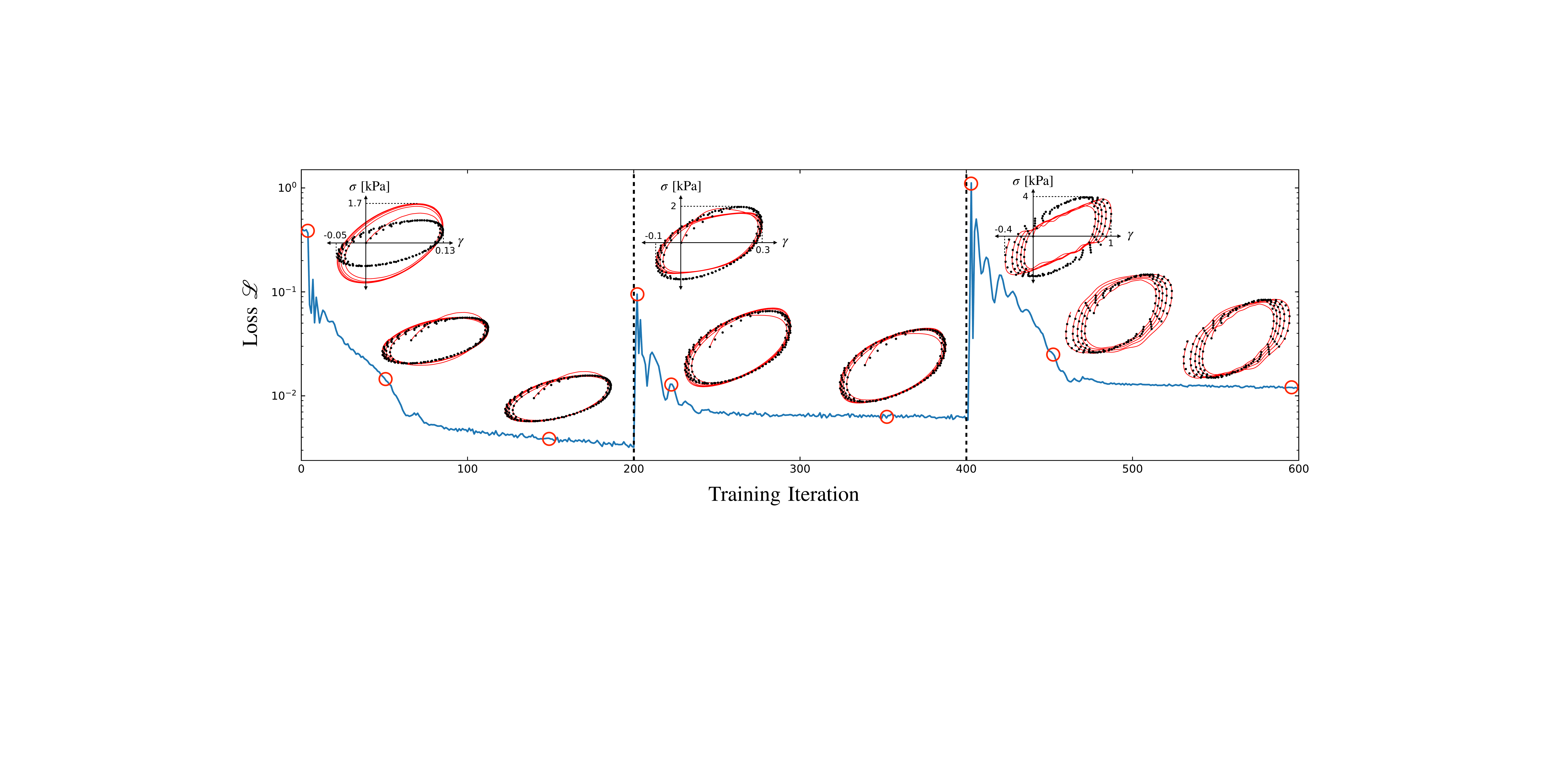}
    \caption{The total loss metric during progression of training. During each epoch (200 iterations), the loss is observed to approximately converge. In the first 200 iterations, the model is trained solely on the $\sigma_0 = 1$ kPa data. Lissajous curves depicting the experimental data (black circles) and model prediction (red line) for this stress amplitude are shown for three select iterations (highlighted with red circles). In iterations 201-400, the model is trained simultaneously on $\sigma_0 = 1$ and $2$ kPa data, with Lissajous curves of the $\sigma_0 = 2$ kPa experiment shown for select iterations. In the final 200 iterations, the model is trained simultaneously on $\sigma_0 = 1$, $2$, and $4$ kPa data, with Lissajous plots of the $\sigma_0 = 4$ kPa experiment shown for select iterations.}
    \label{fig:training}
\end{figure*}

\subsection*{Model Discovery for a Viscoelastic Fluid}

Although the previous example demonstrated many of the salient features of RUDEs, in particular their ability to recover full nonlinear constitutive models from limited data, it did represent a specialized case where the data is exactly described by a simple functional form for $\boldsymbol{F}(\boldsymbol{\sigma},\boldsymbol{\dot{\gamma}})$. For real experimental data, no exact function exists; however, a RUDE is still capable of approximating the data very closely. To illustrate the applicability of RUDEs to real experimental data obtained for viscoelastic fluids, we study a metal-crosslinked polymer hydrogel known to exhibit a Maxwellian linear viscoelastic response \citep{Menyo2013,Song2020}. Using small-amplitude oscillatory shear experiments, we first characterize the linear rheology of this system, and find that it is well-described by a single Maxwell mode ($\tau_2$ = 0) with $G_0 = 37585$ Pa and $\tau_1 = 0.557$ s (Figure S1). We then conducted four stress-controlled LAOS tests with an imposed waveform $\sigma_{12}(t) = \sigma_0 \cos(\omega_0 t)$, at $\omega_0 = 1$ rad/s with $\sigma_0 \in \{1,2,3,4\}$ kPa. As previously discussed, the data-agnostic nature of RUDEs allows stress-controlled experiments to train the model just as well as strain-controlled experiments. The data with $\sigma_0 \in \{1,2,4\}$ kPa were retained for training, while the experiment with $\sigma_0 = 3$ kPa was held out for testing. More information on the experimental methods and materials are given in the Materials and Methods section.

\begin{figure*}
    \centering
    \includegraphics[width=\textwidth]{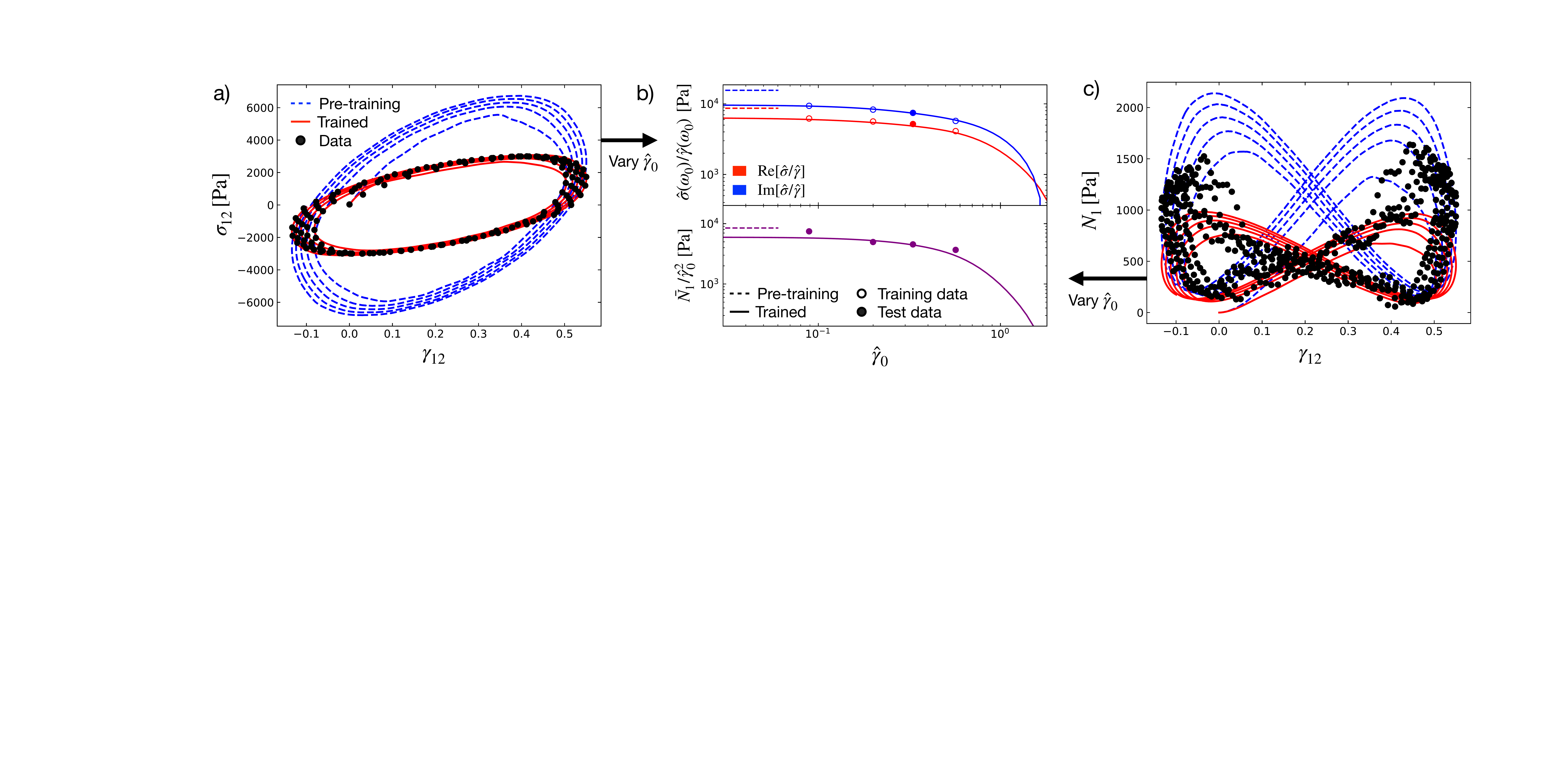}
    \caption{Evaluation of a RUDE trained on experimental oscillatory data for a metal-crosslinked polymer hydrogel. (a) Lissajous curves of the observed shear stress $\sigma_{12}(t)$ versus oscillating shear strain $\gamma_{12}(t)$ from a LAOS experiment with $\sigma_0 = 3$ kPa and $\omega_0 = 1$ rad/s (black circles), compared to the prediction of the trained RUDE (red line) and the initial prediction of the pre-training RUDE (dashed blue line). (b) The measured and predicted evolution of the real (red) and imaginary (blue) components of the first harmonic shear stress response (top), and the measured and predicted average value attained by the first normal stress difference (bottom), in a stress-controlled amplitude sweep, plotted against the first-harmonic strain amplitude $\hat{\gamma}_0 = |\hat{\gamma}(\omega_0)|$. Here, $\hat{\sigma}(\omega_0)$ and $\hat{\gamma}(\omega_0)$ represent the values of the discrete Fourier transforms of the shear stress and shear strain at the fundamental frequency $\omega_0$, respectively. Unfilled circles depict the training data, filled circles depict separate test data, and solid lines depict predictions from the trained RUDE. Predictions of the untrained RUDE are independent of $\hat{\gamma}_0$, and are shown by truncated dashed lines. (c) Lissajous curves of the first normal stress difference in the material that arises from the nonlinear nature of the deformation response with $\sigma_0 = 3$ kPa, compared to the predictions of the trained and pre-training RUDEs.}
    \label{fig:lissajous}
\end{figure*}

Here, we again train the RUDE using only the shear stress and shear strain data obtained from the rheometer, while setting aside the oscillating normal stress data for testing purposes. Figure \ref{fig:training} depicts the training loss as a function of iteration number for the RUDE. We note again that we have applied a continuation approach to training, wherein training data at successively higher amplitudes are incorporated one-by-one after epochs of 200 iterations. Thus, for the first 200 iterations, the RUDE is trained solely on the $\sigma_0 = 1$ kPa data, while for iterations 201-400 it is trained on the $\sigma_0 = 1$ and $2$ kPa data simultaneously, and finally on all three training experiments for the final 200 iterations. One important feature that distinguishes RUDEs from other machine learning modeling approaches is that the RUDE represents a feasible constitutive equation, which respects frame invariance, for every iteration during training. In fully black-box approaches, this is very likely not the case, as feasibility is only achieved near convergence. This salient feature is evidenced by plotting Lissajous curves of the training data and model predictions at select iterations in Figure \ref{fig:training}. Not only do we see a gradual improvement of the model as training progresses, but we see that even when the loss has not converged, the model predictions are stable and represent feasible behavior for a viscoelastic material response (i.e. a phase-delayed response corresponding to an elliptical orbit in the material response plane), although they may not accurately model the actual data.

We may now investigate how the trained RUDE interpolates between data in the training set, and extrapolates to other important features of the nonlinear material response. Figure \ref{fig:lissajous}a) plots the shear stress versus shear strain for the LAOS experiment with $\sigma_0 = 3$ kPa, belonging to the test data set. The trained model accurately captures both the transient and eventual time-periodic steady state observed in the data, a substantial improvement from the pre-training state. Moreover, we note that the model correctly predicts that the oscillations are centered about a non-zero strain by default, a consequence of the linear Maxwellian structure imparted on the RUDE \citep{Hassager2020}. In Figure \ref{fig:lissajous}c), we plot the first normal stress difference $N_1(t;\omega)$ for the same LAOS experiment, and again see that the trained RUDE substantially improves upon the pre-training model in describing the observed first normal stress difference, including both the non-zero mean offset as well as the characteristic amplitude of variations. The experimental normal stress is quite noisy, and not perfectly described by the model predictions; however, it is encouraging that even qualitative features of this response are captured by the trained model, as no normal stress difference data was seen by the model during training. This observation instills confidence that the trained RUDE will provide a reasonable description of the material response in other circumstances that do not fall within the training set, such as in different flow kinematics or different flow protocols. Of course, if this normal stress data was also provided as part of the training set in subsequent epochs, it would also allow for further refinement of the coefficients in the TBNN that forms the `best' RUDE for this unknown viscoelastic material.

Because the trained RUDE represents a materially objective constitutive equation of the Maxwell-Oldroyd type (satisfying what Oldroyd characterized as `rheological invariance' \citep{Oldroyd1984}), we may use it to study aspects of the material response beyond simulating discretely sampled time-series data. For instance, we may predict the first harmonic response in stress-controlled LAOS as a continuous function of the deformation amplitude at any frequency \citep{Hyun2011}. These predictions, shown in Figure \ref{fig:lissajous}b), not only closely agree with the discrete values of training data (unfilled circles) and the test data (filled circle), but provide insight into the nonlinear strain-dependent model predictions at amplitudes beyond those observed experimentally. Indeed, we see that the model predicts a strongly thinning response, and an eventual crossover between the real and imaginary components of the response. Such features are often encountered in nonlinear responses of both experimental data for soft materials as well as the corresponding constitutive equations, and this demonstration shows that RUDEs may be studied in the same way. Similarly, we may predict the evolution in the non-zero mean value of the oscillating first normal stresses developed by the material in nonlinear deformation (denoted $\bar{N}_1$) as a function of deformation amplitude [Figure \ref{fig:lissajous}d)], which again describes the evolution in the data well, and again shows a high degree of strain thinning at deformation amplitudes larger than those observed experimentally.

\subsection*{RUDEs in Computational Fluid Dynamics}

We have noted throughout this work that a trained RUDE represents a true rheological equation of state, in the sense that it relates the history of internal stresses and deformations in a complex fluid without relying on any particular kinematics or experimental protocol. It is materially objetive by construction, and thus satisfied the principles of ``rheological invariance'' as originally elegantly elucidated by Oldroyd \citep{Oldroyd1984}. Therefore, unlike any other previous machine learning approach to complex fluids modeling, it is possible to perform non-trivial, two- or three-dimensional complex fluid dynamics (CFD) simulations using an appropriately trained RUDE. A tool that enables accurate simulations of complex fluids in complicated application-specific flow geometries, while only requiring training data from well-defined and readily laboratory accessible rheometric test deformations, would be of high value for efficient design and engineering of processing operations (e.g. mold-filling, coating operations) involving soft materials. Here, we demonstrate that RUDEs represent a realization of this digital pipeline, by deploying a RUDE trained only using simple shear data in a benchmark CFD problem.

Because RUDEs are structured as tensorial systems of differential equations, they may be directly integrated into existing CFD software that solves the incompressible Cauchy momentum equation:
\begin{equation}
    \rho\frac{D\boldsymbol{u}}{Dt} = -\nabla p + \nabla \cdot \boldsymbol{\sigma}_T + \boldsymbol{f}
\end{equation}
for a fluid of density $\rho$, acted on by body forces $\boldsymbol{f}$, with an isotropic scalar pressure $p(\boldsymbol{x},t)$. Note that $\frac{D}{Dt}$ here represents the material derivative. Here, we simulate a fluid whose deviatoric stress $\boldsymbol{\sigma}_T$ comprises a Newtonian solvent with viscosity $\eta_S$ and a viscoelastic component with stress $\boldsymbol{\sigma}$ that is governed by the RUDE (equation \ref{eq:rude}): $\boldsymbol{\sigma}_T = \boldsymbol{\sigma} + \eta_s\boldsymbol{\dot{\gamma}}$. The `ground truth' fluid is one for which the polymeric stress is given by the Giesekus model with $\alpha = 0.8$, as depicted by the black symbols in Figure \ref{fig:tbnn}, with $\eta_S/\eta_T = 1/9$ where $\eta_T = \eta_S + G_0\tau_1$. As before, the RUDE is trained solely on synthetic LAOS data in simple shear, whose predictions are shown by the red curves in Figure \ref{fig:tbnn}.

\begin{figure*}
    \centering
    \includegraphics[width=\textwidth]{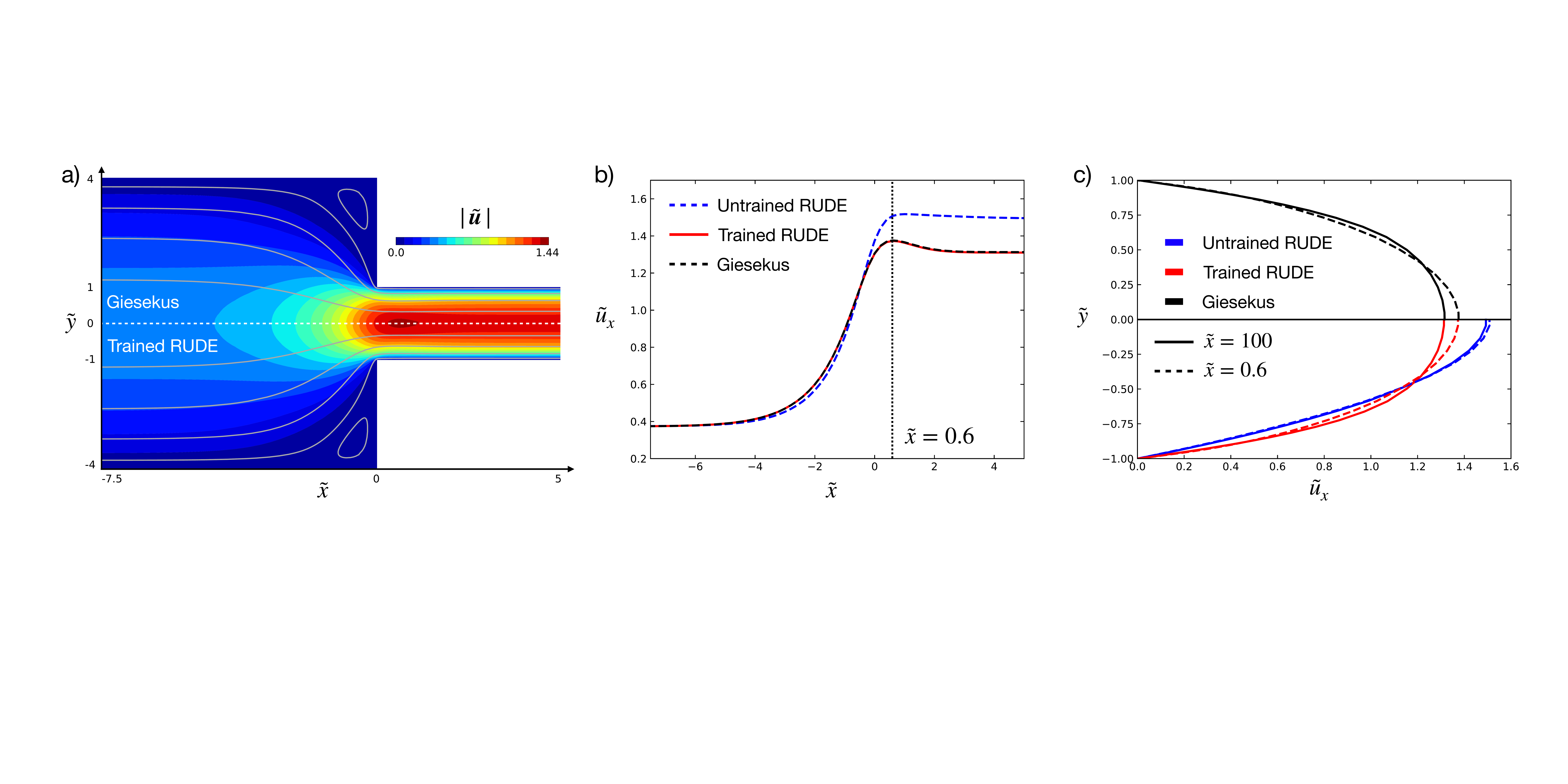}
    \caption{Snapshot of a simulation of flow of a complex fluid into a 4:1 planar contraction with uniform flow of velocity $U$ beginning at $x = -100H$. The abrupt contraction from a channel half-width of $4H$ to a half-width of $H$ occurs at $x = 0$. Data is shown after the flow has reached a steady state ($t = 10 \tau_1$), for $\mathrm{Re} = \rho U H/\eta_T = 0.025$ and $\mathrm{Wi} = \tau_1 U/H = 0.25$. a) The magnitude of the local velocity field around the contraction, in the scaled coordinates $\tilde{\boldsymbol{u}} = \boldsymbol{u}/U$, $\tilde{x} = x/H$, and $\tilde{y} = y/H$. The top half of the panel shows the behavior of the Giesekus model (ground truth), and the bottom half shows the behavior predicted by the trained RUDE. Streamlines are depicted by gray curves. b) The streamwise component of the velocity along the centerline ($y = 0$) in the neighborhood the contraction, with the Giesekus model (ground truth) shown as a black dashed curve, the pre-training RUDE (Oldroyd-B) shown as a blue dashed curve, and the prediction of the trained RUDE shown in red. c) Velocity profiles for the Giesekus model (black), pre-training RUDE (blue), and trained RUDE (red) at the location of maximum velocity overshoot ($\tilde{x} = 0.6$, shown by dashed lines) and in the fully-developed region far downstream of the contraction plane ($\tilde{x} = 100$, shown by solid lines).}
    \label{fig:cfd}
\end{figure*}

Upon integrating the trained RUDE into OpenFOAM \citep{Weller1998}, an open-source CFD software, using the rheoTool plugin \citep{Pimenta2017}, which enables simulations using differential rheological equations of states by the finite volume method, we simulate flow through a 4:1 planar contraction -- a benchmark problem for simulations of viscoelastic fluids that is characteristic of injection molding operations \citep{Alves2003}. The inlet velocity far upstream is set to be uniform with a velocity $U$, with the channel contracting from a width of $8H$ to a width of $2H$ at $x = 0$. Figure \ref{fig:cfd} depicts the simulated flow field at $t/\tau_1 = 10$ for a simulation with $\mathrm{Re} = \rho U H/\eta_T = 0.025$ and $\mathrm{Wi} = \tau_1 U/H = 0.25$, corresponding to an elasticity number \citep{Denn1971} of $\mathrm{El} \equiv \mathrm{Wi}/\mathrm{Re} = 10$ characteristic of extrusion processes for polymer melts \citep{McKinley2005}. More details regarding the simulation setup and solvers are presented in the Materials and Methods section.

We first see from Figure \ref{fig:cfd}a) that the two-dimensional velocity field predicted by the RUDE is visually indistinguishable from the ground-truth predictions of the Giesekus model. This observation is significant -- the RUDE was trained explicitly to mimic the Giesekus model with high accuracy in oscillating simple shearing deformations, but evidently is capable of maintaining this accuracy in a contraction flow which exhibits both shear and extensional characteristics. The close agreement between the ground truth and the trained RUDE is further explored in Figures \ref{fig:cfd}b) and c), which show the streamline velocity $u_x(x, y = 0)$ along the centerline and the velocity profile $v_x(y)$ across the narrow die at two vertical slices ($x = 0.6H$ and $x = 100H$). We again see that the RUDE quantitatively matches the predicted behavior of the Giesekus model, including predicting an overshoot in velocity just beyond the contraction, and a flattening of the velocity profile (due to the shear-rate-dependence of the viscosity) within the narrow die channel as the flow develops. Moreover, it is clear that the model predictions substantially improve as a result of the training, as the pre-training RUDE (i.e. the simplest frame-invariant Oldroyd-B model response) does not accurately predict either of these important characteristic features of this kinematically complex polymer processing operation.

Despite the brevity of this initial investigation of incorporating RUDEs into CFD, the results in this Section reveal the promise of RUDEs as a tool for highly accurate, high-performance scientific simulations of complex fluids. The same trained rheologically invariant model we employed in the planar contraction flow analysis can be used `out-of-the-box' to simulate other two- or three-dimensional flows (e.g. flow around immersed objects), multiphase flows (e.g. die swelling during extrusion) identified by rheoTool \citep{Pimenta2017}, and even in full multiphysics simulations. Due to the selected differential structure of the RUDE, these simulations remain highly scalable, requiring only a small constant multiplicative factor of additional computation in excess of simulations with typical analytic rheological equations of state while providing substantially higher accuracy. Finally, it is worth reemphasizing a key feature that distinguishes RUDEs from other machine learning approaches in complex fluids modeling: that is, the same trained RUDE, with exactly the same set of weights parameterizing the neural network's description of the function $\boldsymbol{F}$, was used to generate \emph{all} of the predictions in Figures \ref{fig:tbnn} and \ref{fig:cfd}. This underscores the portability of RUDEs -- allowing one to train a rheological equation of state on rheometric data that is straightforward to obtain in the laboratory, and then apply the resulting `digital twin' of a complex fluid in flow simulations of industrially relevant process unit operations without needing to obtain new data.

\section*{Discussion}

The previous section has demonstrated that RUDEs represent capable digital tools for discovering materially objective constitutive equations, and that these models make highly accurate predictions when interpolating between training data conditions. We have also emphasized the versatility of RUDEs, in that trained RUDEs make transferable predictions of observables not seen during training, and are portable to flows outside of those used to generate training data, including multidimensional flows that must be investigated by CFD. This versatility extends to different training scenarios, in which we may simultaneously train a single RUDE on multiple types of data, whether a combination of shear and normal stress data, oscillatory tests with a range of deformation amplitudes and frequencies, or stress growth data following flow inception. Flexibility in assimilating diverse data was a motivating principle behind the development of the RUDE framework, inspired by purely physical or empirical constitutive equations that are often parameterized by many types of experiments and observables. Therefore, it is worth briefly considering other ways in which this flexibility may be exploited.

In our development of RUDEs, we have limited ourselves to frame-invariant differential models of the Maxwell-Oldroyd type \citep{Lennon2021} with a single relaxation mode as the foundation. This limitation is not necessary, however. One can readily add additional physics-informed terms to the Oldroyd formulation if so desired, to provide an initialization for the RUDE that is closer to the expected behavior. One can also incorporate more complex phenomenology into this simple viscoelastic framework, such as kinetic evolution equations for internal microstructural variables that give rise to thixotropy, plasticity, or a combination of the two \citep{Dimitriou2014}. Extending the framework to multiple relaxation modes (a common feature of many commercial materials) is also straightforward. Moreover, due to the inherent parallelism of the resulting `multi-mode' models \citep{Bird2016}, in combination with task-level parallelism across experiments in the loss function (equation \ref{eq:loss}), one may distribute computation of the gradient across multiple threads or processors, providing a highly scalable framework for model training. This scalability may open an avenue to big data applications of RUDEs, leading to digital twins of complex fluids that are highly accurate across a very wide variety of experimental conditions.

\section*{Conclusions}

We have introduced the rheological universal differential equation (RUDE), a learnable constitutive model framework that may be trained and subsequently used to predict any experimental observable in arbitrary flows, and which directly incorporates physical or empirical knowledge into its structure. Due to the embedded tensor basis neural network, RUDEs quickly learn simple yet accurate and highly general models for accurately describing the provided training data. The formulation of RUDEs as differential viscoelastic equations of state facilitates their direct integration into existing computational fluid dynamics tools. Therefore, RUDEs represent a new avenue for the rapid and accurate modeling of complex fluids in industrially relevant processes. They also represent a highly customizable framework, that may be imparted with new physical or empirical insight based on the specific material system under study or the relevant processing application. We hope that the presentation of such a customizable yet robust method for data-driven modeling of soft materials does for rheology what previous machine learning frameworks have done for so many other fields, allowing the automated development of digital twins for rheologically complex fluids that facilitate rapid computational design and optimization of real-world flows and processing operations.

\section*{Materials and Methods}

\subsection*{Tensor Basis Neural Network}

The Tensor Basis Neural Network employed in this work is a linear basis expansion presented in equation \ref{eq:tbnn}. The tensors $\boldsymbol{T}_n$ are derived by repeated inner products of the stress and deformation rate tensors $\boldsymbol{\sigma}$, $\boldsymbol{\dot{\gamma}}$, with the identity tensor $\boldsymbol{\delta}$. The set of $\boldsymbol{T}_n$ is closed by application of the Cayley-Hamilton theorem, leaving the following set of nine tensors:
\begin{equation}
    \boldsymbol{T}_n = \begin{cases}\boldsymbol{\delta}, ~\boldsymbol{\sigma}, ~\boldsymbol{\dot{\gamma}}, ~\boldsymbol{\sigma}\cdot\boldsymbol{\sigma}, ~\boldsymbol{\dot{\gamma}}\cdot\boldsymbol{\dot{\gamma}}, ~\boldsymbol{\sigma}\cdot\boldsymbol{\dot{\gamma}} + \boldsymbol{\dot{\gamma}}\cdot\boldsymbol{\sigma}, \\
    \boldsymbol{\sigma}\cdot\boldsymbol{\sigma}\cdot\boldsymbol{\dot{\gamma}} + \boldsymbol{\dot{\gamma}}\cdot\boldsymbol{\sigma}\cdot\boldsymbol{\sigma}, ~\boldsymbol{\sigma}\cdot\boldsymbol{\dot{\gamma}}\cdot\boldsymbol{\dot{\gamma}} + \boldsymbol{\dot{\gamma}}\cdot\boldsymbol{\dot{\gamma}}\cdot\boldsymbol{\sigma}, \\
    \boldsymbol{\sigma}\cdot\boldsymbol{\sigma}\cdot\boldsymbol{\dot{\gamma}}\cdot\boldsymbol{\dot{\gamma}} + \boldsymbol{\dot{\gamma}}\cdot\boldsymbol{\dot{\gamma}}\cdot\boldsymbol{\sigma}\cdot\boldsymbol{\sigma}\end{cases}.
\end{equation}
The coefficients in the basis expansion are the outputs of a fully-connected feedforward neural network, whose inputs are the independent invariants of the tensors $\boldsymbol{T}_n$. It is possible to express these invariants in terms of the first invariant (i.e. the trace) of these and higher-order tensors, leaving the following closed set of tensor invariants (assuming that the viscoelastic material is incompressible, such that the rate-of-strain tensor is traceless, i.e. $\mathrm{tr}(\boldsymbol{\dot{\gamma}}) = 0$):
\begin{equation}
    \lambda_n = \begin{cases}
    \mathrm{tr}(\boldsymbol{\sigma}), ~\mathrm{tr}(\boldsymbol{\sigma}\cdot\boldsymbol{\sigma}), ~\mathrm{tr}(\boldsymbol{\dot{\gamma}}\cdot\boldsymbol{\dot{\gamma}}), ~\mathrm{tr}(\boldsymbol{\sigma}\cdot\boldsymbol{\sigma}\cdot\boldsymbol{\sigma}), \\
    \mathrm{tr}(\boldsymbol{\dot{\gamma}}\cdot\boldsymbol{\dot{\gamma}}\cdot\boldsymbol{\dot{\gamma}}), ~\mathrm{tr}(\boldsymbol{\sigma}\cdot\boldsymbol{\dot{\gamma}}), ~\mathrm{tr}(\boldsymbol{\sigma}\cdot\boldsymbol{\sigma}\cdot\boldsymbol{\dot{\gamma}}), \\
    \mathrm{tr}(\boldsymbol{\sigma}\cdot\boldsymbol{\dot{\gamma}}\cdot\boldsymbol{\dot{\gamma}}), ~\mathrm{tr}(\boldsymbol{\sigma}\cdot\boldsymbol{\sigma}\cdot\boldsymbol{\dot{\gamma}}\cdot\boldsymbol{\dot{\gamma}})
    \end{cases}.
\end{equation}
The neural network employed in this work has two hidden layers, each with 32 neurons. The internal activations of the network are hyperbolic tangent functions, selected so that the neural network is everywhere differentiable with respect to the inputs. The network was trained for 200 iterations per input experiment using the AMSGrad optimizer, with an $L^1$ penalty (weighted by 0.01) imposed during training.

\subsection*{Hydrogel Synthesis}

Synthesis of nitrocatechol-functionalized 4-arm poly(ethylene glycol) (4nPEG) follows the protocol in reference \cite{Song2020}. Hydrogels were prepared by first mixing a solution of 4nPEG with FeCl$_3$ in a stoichiometric ratio of 3 nitrocatechol to 1 Fe$^{3+}$.  The mixture was then buffered with bicine (pH = 8.5) to induce nitrocatechol deprotonation, leading to metal-coordination cross-linking and gelation. The final polymer concentration in the hydrogel was 10 wt. \%, and the final buffer concentration in the hydrogel was 0.2 M.

\subsection*{Experimental Methods}

All experiments presented in this work were conducted using a DHR-3 Discovery Hybrid Rheometer from TA Instruments, with an 8mm parallel plate geometry and bottom Peltier plate maintaining a set temperature of 25$^\circ$C. The small amplitude frequency sweep was conducted in stress control with a stress amplitude of $\sigma_0 = 5$ Pa, and data was converted to the linear complex modulus using the TRIOS software v5.2.2. All large amplitude tests were run in stress control with a sampling rate of 488 points per second and for a duration of five periods with respect to the imposed frequency of oscillation, with time series data output by TRIOS. 

\subsection*{CFD Simulation Methods}

The CFD simulation presented in this work was performed using OpenFOAM \cite{Weller1998} with the rheoTool package \cite{Pimenta2017}. The simulation mesh employed in the 4:1 planar contraction had an inlet face at $x = -100H$ over the interval $y \in [-4H,4H]$, top and bottom walls at $y = \pm 4H$ for $x < 0$ and $y = \pm H$ for $x \geq 0$, vertical walls at $x = 0$ over the intervals $y \in [H,4H]$ and $y \in [-4H,-H]$, and an outlet face at $x = 100H$ over the interval $y \in [-H,H]$. The channel was discretized over a Cartesian grid. The velocity boundary conditions were specified as a zero-gradient condition at the outlet; zero-velocity conditions (i.e. no-slip and no-penetration) at the top, bottom, and vertical walls; and the following time-varying uniform velocity profile at the inlet:
\begin{equation}
    \frac{\boldsymbol{u}(t)}{U} = \begin{cases}\frac{1 - \cos(\pi t/\tau_1)}{2}\mathbf{e}_x & t < \tau_1 \\
    \mathbf{e}_x & t \geq \tau_1\end{cases},
\end{equation}
such that $u_x(t)$ increases smoothly from $0$ at $t = 0$ to $U$ at $t = \tau_1$. The pressure boundary conditions were set to a zero-gradient condition at the inlet and walls, and a fixed value of zero at the outlet. The deviatoric stress boundary conditions were set to a zero-gradient condition at the outlet, a fixed-value of $\boldsymbol{0}$ condition at the inlet, and a linear extrapolation to zero at the walls (preferred over a zero-gradient condition due to the higher order of computational accuracy \cite{Pimenta2017}).

Differential operators in space were evaluated using the finite volume method with linear interpolation and Gaussian integration, with integration in time achieved by the implicit Euler method with a fixed time step of $(2 \times 10^{-4})\tau_1$. At each time step, the velocity, pressure, and stress fields were computed using the Semi-Implicit Method for Pressure Linked Equations (SIMPLE), using the preconditioned conjugate gradient method (with a diagonal incomplete-Cholesky preconditioner) to solve for the pressure, and the preconditioned bi-conjugate gradient method (with a diagonal incomplete-LU preconditioner) to solve for the velocity and stress.

\section*{Acknowledgements}

The authors thank Jake Song for providing the hydrogel system and assistance with experimentation. K.R.L. was supported by the U.S. Department of Energy Compu- tational Science Graduate Fellowship program under Grant No. DE- SC0020347.

\bibliographystyle{unsrtnat}
\bibliography{references}  

\begin{thebibliography}{40}
\providecommand{\natexlab}[1]{#1}
\providecommand{\url}[1]{\texttt{#1}}
\expandafter\ifx\csname urlstyle\endcsname\relax
  \providecommand{\doi}[1]{doi: #1}\else
  \providecommand{\doi}{doi: \begingroup \urlstyle{rm}\Url}\fi

\bibitem[Carleo et~al.(2019)Carleo, Cirac, Cranmer, Daudet, Schuld, Tishby,
  Vogt-Maranto, and Zdeborov\'a]{Carleo2019}
Giuseppe Carleo, Ignacio Cirac, Kyle Cranmer, Laurent Daudet, Maria Schuld,
  Naftali Tishby, Leslie Vogt-Maranto, and Lenka Zdeborov\'a.
\newblock Machine learning and the physical sciences.
\newblock \emph{Rev. Mod. Phys.}, 91:\penalty0 045002, Dec 2019.
\newblock \doi{10.1103/RevModPhys.91.045002}.
\newblock URL \url{https://link.aps.org/doi/10.1103/RevModPhys.91.045002}.

\bibitem[Goh et~al.(2017)Goh, Hodas, and Vishnu]{Goh2017}
Garrett~B. Goh, Nathan~O. Hodas, and Abhinav Vishnu.
\newblock Deep learning for computational chemistry.
\newblock \emph{Journal of Computational Chemistry}, 38\penalty0 (16):\penalty0
  1291--1307, 2017.
\newblock \doi{https://doi.org/10.1002/jcc.24764}.
\newblock URL \url{https://onlinelibrary.wiley.com/doi/abs/10.1002/jcc.24764}.

\bibitem[Park et~al.(2018)Park, Took, and Seong]{Park2018}
Cheolsoo Park, Clive~Cheong Took, and Joon-Kyung Seong.
\newblock {Machine learning in biomedical engineering}.
\newblock \emph{Biomedical Engineering Letters}, 8\penalty0 (1):\penalty0 1--3,
  2018.
\newblock ISSN 2093-985X.
\newblock \doi{10.1007/s13534-018-0058-3}.
\newblock URL \url{https://doi.org/10.1007/s13534-018-0058-3}.

\bibitem[Radovic et~al.(2018)Radovic, Williams, Rousseau, Kagan, Bonacorsi,
  Himmel, Aurisano, Terao, and Wongjirad]{Radovic2018}
Alexander Radovic, Mike Williams, David Rousseau, Michael Kagan, Daniele
  Bonacorsi, Alexander Himmel, Adam Aurisano, Kazuhiro Terao, and Taritree
  Wongjirad.
\newblock {Machine learning at the energy and intensity frontiers of particle
  physics}.
\newblock \emph{Nature}, 560\penalty0 (7716):\penalty0 41--48, 2018.
\newblock ISSN 1476-4687.
\newblock \doi{10.1038/s41586-018-0361-2}.
\newblock URL \url{https://doi.org/10.1038/s41586-018-0361-2}.

\bibitem[Rackauckas et~al.(2020)Rackauckas, Ma, Martensen, Warner, Zubov,
  Supekar, Skinner, Ramadhan, and Edelman]{Rackauckas2020}
Christopher Rackauckas, Yingbo Ma, Julius Martensen, Collin Warner, Kirill
  Zubov, Rohit Supekar, Dominic Skinner, Ali Ramadhan, and Alan Edelman.
\newblock Universal differential equations for scientific machine learning,
  2020.
\newblock URL \url{https://arxiv.org/abs/2001.04385}.
\newblock arXiv:2001.04385.

\bibitem[Ellis et~al.(2021)Ellis, Fiedler, Popoola, Modine, Stephens, Thompson,
  Cangi, and Rajamanickam]{Ellis2021}
J.~A. Ellis, L.~Fiedler, G.~A. Popoola, N.~A. Modine, J.~A. Stephens, A.~P.
  Thompson, A.~Cangi, and S.~Rajamanickam.
\newblock Accelerating finite-temperature kohn-sham density functional theory
  with deep neural networks.
\newblock \emph{Phys. Rev. B}, 104:\penalty0 035120, Jul 2021.
\newblock \doi{10.1103/PhysRevB.104.035120}.
\newblock URL \url{https://link.aps.org/doi/10.1103/PhysRevB.104.035120}.

\bibitem[Ling et~al.(2016)Ling, Kurzawski, and Templeton]{Ling2016}
Julia Ling, Andrew Kurzawski, and Jeremy Templeton.
\newblock {Reynolds averaged turbulence modelling using deep neural networks
  with embedded invariance}.
\newblock \emph{Journal of Fluid Mechanics}, 807:\penalty0 155--166, 2016.
\newblock ISSN 0022-1120.
\newblock \doi{DOI: 10.1017/jfm.2016.615}.
\newblock URL
  \url{https://www.cambridge.org/core/article/reynolds-averaged-turbulence-modelling-using-deep-neural-networks-with-embedded-invariance/0B280EEE89C74A7BF651C422F8FBD1EB}.

\bibitem[W.~Tschoegl(1989)]{Tschoegl1989}
Nicholas W.~Tschoegl.
\newblock \emph{The Phenomenological Theory of Linear Viscoelastic Behavior: An
  Introduction}.
\newblock Springer-Verlag, Berlin, 01 1989.
\newblock \doi{10.1007/978-3-642-73602-5_3}.

\bibitem[Bird et~al.(1987)Bird, Armstrong, and Hassager]{Bird1987}
R.B Bird, R.C. Armstrong, and Ole Hassager.
\newblock \emph{Dynamics of Polymeric Liquids, Volume 1: Fluid Mechanics}.
\newblock John Wiley \& Sons, Inc., 05 1987.
\newblock \doi{10.1002/pol.1978.130160210}.

\bibitem[Svozil et~al.(1997)Svozil, Kvasnicka, and Pospichal]{Svozil1997}
Daniel Svozil, Vladimir Kvasnicka, and Jiri Pospichal.
\newblock {Introduction to multi-layer feed-forward neural networks}.
\newblock \emph{Chemometrics and Intelligent Laboratory Systems}, 39\penalty0
  (1):\penalty0 43--62, 1997.
\newblock ISSN 0169-7439.
\newblock \doi{https://doi.org/10.1016/S0169-7439(97)00061-0}.
\newblock URL
  \url{https://www.sciencedirect.com/science/article/pii/S0169743997000610}.

\bibitem[Mahmoudabadbozchelou et~al.(2021)Mahmoudabadbozchelou, Caggioni,
  Shahsavari, Hartt, Em~Karniadakis, and Jamali]{Mahmoudabadbozchelou2021a}
Mohammadamin Mahmoudabadbozchelou, Marco Caggioni, Setareh Shahsavari,
  William~H. Hartt, George Em~Karniadakis, and Safa Jamali.
\newblock Data-driven physics-informed constitutive metamodeling of complex
  fluids: A multifidelity neural network (mfnn) framework.
\newblock \emph{Journal of Rheology}, 65\penalty0 (2):\penalty0 179--198, 2021.
\newblock \doi{10.1122/8.0000138}.
\newblock URL \url{https://doi.org/10.1122/8.0000138}.

\bibitem[Mahmoudabadbozchelou and Jamali(2021)]{Mahmoudabadbozchelou2021b}
Mohammadamin Mahmoudabadbozchelou and Safa Jamali.
\newblock {Rheology-Informed Neural Networks (RhINNs) for forward and inverse
  metamodelling of complex fluids}.
\newblock \emph{Scientific Reports}, 11\penalty0 (1):\penalty0 12015, 2021.
\newblock ISSN 2045-2322.
\newblock \doi{10.1038/s41598-021-91518-3}.
\newblock URL \url{https://doi.org/10.1038/s41598-021-91518-3}.

\bibitem[Mahmoudabadbozchelou et~al.(2022{\natexlab{a}})Mahmoudabadbozchelou,
  Karniadakis, and Jamali]{Mahmoudabadbozchelou2022a}
Mohammadamin Mahmoudabadbozchelou, George~Em. Karniadakis, and Safa Jamali.
\newblock {nn-PINNs: Non-Newtonian physics-informed neural networks for complex
  fluid modeling}.
\newblock \emph{Soft Matter}, 18\penalty0 (1):\penalty0 172--185,
  2022{\natexlab{a}}.
\newblock ISSN 1744-683X.
\newblock \doi{10.1039/D1SM01298C}.
\newblock URL \url{http://dx.doi.org/10.1039/D1SM01298C}.

\bibitem[Raissi et~al.(2019)Raissi, Perdikaris, and Karniadakis]{Raissi2019}
M~Raissi, P~Perdikaris, and G~E Karniadakis.
\newblock {Physics-informed neural networks: A deep learning framework for
  solving forward and inverse problems involving nonlinear partial differential
  equations}.
\newblock \emph{Journal of Computational Physics}, 378:\penalty0 686--707,
  2019.
\newblock ISSN 0021-9991.
\newblock \doi{https://doi.org/10.1016/j.jcp.2018.10.045}.
\newblock URL
  \url{https://www.sciencedirect.com/science/article/pii/S0021999118307125}.

\bibitem[Saadat et~al.(2022)Saadat, Mahmoudabadbozchelou, and
  Jamali]{Saadat2022}
Milad Saadat, Mohammadamin Mahmoudabadbozchelou, and Safa Jamali.
\newblock {Data-driven selection of constitutive models via rheology-informed
  neural networks (RhINNs)}.
\newblock \emph{Rheologica Acta}, 2022.
\newblock ISSN 1435-1528.
\newblock \doi{10.1007/s00397-022-01357-w}.
\newblock URL \url{https://doi.org/10.1007/s00397-022-01357-w}.

\bibitem[Thakur et~al.(2022)Thakur, Raissi, and Ardekani]{Thakur2022}
Sukirt Thakur, Maziar Raissi, and Arezoo~M. Ardekani.
\newblock Viscoelasticnet: A physics informed neural network framework for
  stress discovery and model selection, 2022.
\newblock URL \url{https://arxiv.org/abs/2209.06972}.
\newblock arXiv:2201.08363.

\bibitem[Mahmoudabadbozchelou et~al.(2022{\natexlab{b}})Mahmoudabadbozchelou,
  Kamani, Rogers, and Jamali]{Mahmoudabadbozchelou2022b}
Mohammadamin Mahmoudabadbozchelou, Krutarth~M. Kamani, Simon~A. Rogers, and
  Safa Jamali.
\newblock Digital rheometer twins: Learning the hidden rheology of complex
  fluids through rheology-informed graph neural networks.
\newblock \emph{Proceedings of the National Academy of Sciences}, 119\penalty0
  (20):\penalty0 e2202234119, 2022{\natexlab{b}}.
\newblock \doi{10.1073/pnas.2202234119}.
\newblock URL \url{https://www.pnas.org/doi/abs/10.1073/pnas.2202234119}.

\bibitem[Green and Rivlin(1957)]{Green1957}
A~E Green and R~S Rivlin.
\newblock {The mechanics of non-linear materials with memory}.
\newblock \emph{Arch. Ration. Mech. Anal.}, 1:\penalty0 1, 1957.
\newblock \doi{10.1007/BF00297992}.
\newblock URL \url{https://doi.org/10.1007/BF00297992}.

\bibitem[Green and Rivlin(1959)]{Green1959}
A~E Green and R~S Rivlin.
\newblock {The mechanics of non-linear materials with memory}.
\newblock \emph{Arch. Ration. Mech. Anal.}, 4:\penalty0 387, 1959.
\newblock \doi{10.1007/BF00281398}.
\newblock URL \url{https://doi.org/10.1007/BF00281398}.

\bibitem[Oldroyd(1984)]{Oldroyd1984}
J~G Oldroyd.
\newblock {An approach to non-newtonian fluid mechanics}.
\newblock \emph{Journal of Non-Newtonian Fluid Mechanics}, 14:\penalty0 9--46,
  1984.
\newblock ISSN 0377-0257.
\newblock \doi{https://doi.org/10.1016/0377-0257(84)80035-X}.
\newblock URL
  \url{https://www.sciencedirect.com/science/article/pii/037702578480035X}.

\bibitem[Oldroyd and Wilson(1950)]{Oldroyd1950}
J~G Oldroyd and Alan~Herries Wilson.
\newblock {On the formulation of rheological equations of state}.
\newblock \emph{Proceedings of the Royal Society of London. Series A.
  Mathematical and Physical Sciences}, 200\penalty0 (1063):\penalty0 523--541,
  feb 1950.
\newblock \doi{10.1098/rspa.1950.0035}.
\newblock URL \url{https://doi.org/10.1098/rspa.1950.0035}.

\bibitem[Bampi and Morro(1980)]{Bampi1980}
F~Bampi and A~Morro.
\newblock {Objectivity and objective time derivatives in continuum physics}.
\newblock \emph{Foundations of Physics}, 10\penalty0 (11):\penalty0 905--920,
  1980.
\newblock ISSN 1572-9516.
\newblock \doi{10.1007/BF00708688}.
\newblock URL \url{https://doi.org/10.1007/BF00708688}.

\bibitem[Spencer and Rivlin(1958)]{Spencer1958}
A~J~M Spencer and R~S Rivlin.
\newblock {The theory of matrix polynomials and its application to the
  mechanics of isotropic continua}.
\newblock \emph{Archive for Rational Mechanics and Analysis}, 2\penalty0
  (1):\penalty0 309--336, 1958.
\newblock ISSN 1432-0673.
\newblock \doi{10.1007/BF00277933}.
\newblock URL \url{https://doi.org/10.1007/BF00277933}.

\bibitem[Spencer and Rivlin(1959)]{Spencer1959}
A.~J.M. Spencer and R.~S. Rivlin.
\newblock {Further results in the theory of matrix polynomials}.
\newblock \emph{Archive for Rational Mechanics and Analysis}, 4\penalty0 (1
  Supplement):\penalty0 214--230, 1959.
\newblock ISSN 00039527.
\newblock \doi{10.1007/BF00281388}.

\bibitem[Cao et~al.(2003)Cao, Li, Petzold, and Serban]{Cao2003}
Yang Cao, Shengtai Li, Linda Petzold, and Radu Serban.
\newblock Adjoint sensitivity analysis for differential-algebraic equations:
  The adjoint dae system and its numerical solution.
\newblock \emph{SIAM Journal on Scientific Computing}, 24\penalty0
  (3):\penalty0 1076--1089, 2003.
\newblock \doi{10.1137/S1064827501380630}.
\newblock URL \url{https://doi.org/10.1137/S1064827501380630}.

\bibitem[Rackauckas and Nie(2017)]{Rackauckas2017}
Christopher Rackauckas and Qing Nie.
\newblock Differential{E}quations.jl -- a performant and feature-rich ecosystem
  for solving differential equations in {J}ulia.
\newblock \emph{Journal of Open Research Software}, 5\penalty0 (1):\penalty0
  15, 2017.
\newblock \doi{10.5334/jors.151}.
\newblock URL \url{http://doi.org/10.5334/jors.151}.

\bibitem[Rackauckas et~al.(2019)Rackauckas, Innes, Ma, Bettencourt, White, and
  Dixit]{Rackauckas2019}
Chris Rackauckas, Mike Innes, Yingbo Ma, Jesse Bettencourt, Lyndon White, and
  Vaibhav Dixit.
\newblock Diff{E}q{F}lux.jl -- a {J}ulia library for neural differential
  equations.
\newblock \emph{arXiv preprint arXiv:1902.02376}, 2019.
\newblock \doi{10.48550/arXiv.1902.02376}.
\newblock URL \url{https://doi.org/10.48550/arXiv.1902.02376}.

\bibitem[Lennon et~al.(2021)Lennon, McKinley, and Swan]{Lennon2021}
Kyle~R Lennon, Gareth~H McKinley, and James~W Swan.
\newblock {The medium amplitude response of nonlinear Maxwell–Oldroyd type
  models in simple shear}.
\newblock \emph{Journal of Non-Newtonian Fluid Mechanics}, 295:\penalty0
  104601, 2021.
\newblock ISSN 0377-0257.
\newblock \doi{https://doi.org/10.1016/j.jnnfm.2021.104601}.
\newblock URL
  \url{https://www.sciencedirect.com/science/article/pii/S037702572100104X}.

\bibitem[Giesekus(1982)]{Giesekus1982}
H~Giesekus.
\newblock {A simple constitutive equation for polymer fluids based on the
  concept of deformation-dependent tensorial mobility}.
\newblock \emph{Journal of Non-Newtonian Fluid Mechanics}, 11\penalty0
  (1):\penalty0 69--109, 1982.
\newblock ISSN 0377-0257.
\newblock \doi{https://doi.org/10.1016/0377-0257(82)85016-7}.
\newblock URL
  \url{http://www.sciencedirect.com/science/article/pii/0377025782850167}.

\bibitem[Hyun et~al.(2011)Hyun, Wilhelm, Klein, Cho, Nam, Ahn, Lee, Ewoldt, and
  McKinley]{Hyun2011}
K~Hyun, M~Wilhelm, C~O Klein, K~S Cho, J~G Nam, K~H Ahn, S~J Lee, R~H Ewoldt,
  and G~H McKinley.
\newblock {A review of nonlinear oscillatory shear tests: Analysis and
  application of large amplitude oscillatory shear (LAOS)}.
\newblock \emph{Prog. Polym. Sci.}, 36\penalty0 (12):\penalty0 1697, 2011.
\newblock \doi{10.1016/j.progpolymsci.2011.02.002}.
\newblock URL \url{https://doi.org/10.1016/j.progpolymsci.2011.02.002}.

\bibitem[Menyo et~al.(2013)Menyo, Hawker, and Waite]{Menyo2013}
Matthew~S Menyo, Craig~J Hawker, and J~Herbert Waite.
\newblock {Versatile tuning of supramolecular hydrogels through metal
  complexation of oxidation-resistant catechol-inspired ligands}.
\newblock \emph{Soft Matter}, 9\penalty0 (43):\penalty0 10314--10323, 2013.
\newblock ISSN 1744-683X.
\newblock \doi{10.1039/C3SM51824H}.
\newblock URL \url{http://dx.doi.org/10.1039/C3SM51824H}.

\bibitem[Song et~al.(2020)Song, Rizvi, Lynch, Ilavsky, Mankus, Tracy, McKinley,
  and Holten-Andersen]{Song2020}
Jake Song, Mehedi~H Rizvi, Brian~B Lynch, Jan Ilavsky, David Mankus, Joseph~B
  Tracy, Gareth~H McKinley, and Niels Holten-Andersen.
\newblock {Programmable Anisotropy and Percolation in Supramolecular Patchy
  Particle Gels}.
\newblock \emph{ACS Nano}, 14\penalty0 (12):\penalty0 17018--17027, dec 2020.
\newblock ISSN 1936-0851.
\newblock \doi{10.1021/acsnano.0c06389}.
\newblock URL \url{https://doi.org/10.1021/acsnano.0c06389}.

\bibitem[Hassager(2020)]{Hassager2020}
Ole Hassager.
\newblock Stress-controlled oscillatory flow initiated at time zero: A linear
  viscoelastic analysis.
\newblock \emph{Journal of Rheology}, 64\penalty0 (3):\penalty0 545--550, 2020.
\newblock \doi{10.1122/1.5127827}.
\newblock URL \url{https://doi.org/10.1122/1.5127827}.

\bibitem[Weller et~al.(1998)Weller, Tabor, Jasak, and Fureby]{Weller1998}
H~G Weller, G~Tabor, H~Jasak, and C~Fureby.
\newblock {A tensorial approach to computational continuum mechanics using
  object-oriented techniques}.
\newblock \emph{Computers in Physics}, 12\penalty0 (6):\penalty0 620--631, nov
  1998.
\newblock ISSN 0894-1866.
\newblock \doi{10.1063/1.168744}.
\newblock URL \url{https://aip.scitation.org/doi/abs/10.1063/1.168744}.

\bibitem[Pimenta and Alves(2017)]{Pimenta2017}
F~Pimenta and M~A Alves.
\newblock {Stabilization of an open-source finite-volume solver for
  viscoelastic fluid flows}.
\newblock \emph{Journal of Non-Newtonian Fluid Mechanics}, 239:\penalty0
  85--104, 2017.
\newblock ISSN 0377-0257.
\newblock \doi{https://doi.org/10.1016/j.jnnfm.2016.12.002}.
\newblock URL
  \url{https://www.sciencedirect.com/science/article/pii/S0377025716303329}.

\bibitem[Alves et~al.(2003)Alves, Oliveira, and Pinho]{Alves2003}
Manuel~A Alves, Paulo~J Oliveira, and Fernando~T Pinho.
\newblock {Benchmark solutions for the flow of Oldroyd-B and PTT fluids in
  planar contractions}.
\newblock \emph{Journal of Non-Newtonian Fluid Mechanics}, 110\penalty0
  (1):\penalty0 45--75, 2003.
\newblock ISSN 0377-0257.
\newblock \doi{https://doi.org/10.1016/S0377-0257(02)00191-X}.
\newblock URL
  \url{https://www.sciencedirect.com/science/article/pii/S037702570200191X}.

\bibitem[Denn and Porteous(1971)]{Denn1971}
M~M Denn and K~C Porteous.
\newblock {Elastic effects in flow of viscoelastic liquids}.
\newblock \emph{The Chemical Engineering Journal}, 2\penalty0 (4):\penalty0
  280--286, 1971.
\newblock ISSN 0300-9467.
\newblock \doi{https://doi.org/10.1016/0300-9467(71)85007-4}.
\newblock URL
  \url{https://www.sciencedirect.com/science/article/pii/0300946771850074}.

\bibitem[McKinley(2005)]{McKinley2005}
G.H. McKinley.
\newblock Dimensionless groups for understanding free surface flows of complex
  fluids.
\newblock Technical Report 05-P-05, Hatsopoulos Microfluids Laboratory, MIT,
  July 2005.
\newblock URL
  \url{https://dspace.mit.edu/bitstream/handle/1721.1/18086/05-P-05.pdf?sequence=1}.

\bibitem[Dimitriou and McKinley(2014)]{Dimitriou2014}
Christopher~J Dimitriou and Gareth~H McKinley.
\newblock {A comprehensive constitutive law for waxy crude oil: a thixotropic
  yield stress fluid}.
\newblock \emph{Soft Matter}, 10\penalty0 (35):\penalty0 6619--6644, 2014.
\newblock ISSN 1744-683X.
\newblock \doi{10.1039/C4SM00578C}.
\newblock URL \url{http://dx.doi.org/10.1039/C4SM00578C}.

\bibitem[Bird and Giacomin(2016)]{Bird2016}
R.B. Bird and A.J. Giacomin.
\newblock Polymer fluid dynamics: Continuum and molecular approaches.
\newblock \emph{Annual Review of Chemical and Biomolecular Engineering},
  7\penalty0 (1):\penalty0 479--507, 2016.
\newblock \doi{10.1146/annurev-chembioeng-080615-034536}.
\newblock URL \url{https://doi.org/10.1146/annurev-chembioeng-080615-034536}.
\newblock PMID: 27276553.

\end{thebibliography}






\end{document}